\documentclass[12pt,preprint]{emulateapj}
\usepackage{graphicx}

\shorttitle{The HD 10180 System}
\shortauthors{Stephen R. Kane \& Dawn M. Gelino}
\slugcomment{Submitted for publication in the Astrophysical Journal}

\begin{document}

\title{On the Inclination and Habitability of the HD 10180 System}
\author{
  Stephen R. Kane\altaffilmark{1},
  Dawn M. Gelino\altaffilmark{2}
}
\email{skane@sfsu.edu}
\altaffiltext{1}{Department of Physics \& Astronomy, San Francisco
  State University, 1600 Holloway Avenue, San Francisco, CA 94132,
  USA}
\altaffiltext{2}{NASA Exoplanet Science Institute, Caltech, MS 100-22,
  770 South Wilson Avenue, Pasadena, CA 91125, USA}

%%%%%%%%%%%%%%%%%%%%%%%%%%%%%%%%%%%%%%%%%%%%%%%%%%%%%%%%%%%%%%%%%%%%

\begin{abstract}

There are numerous multi-planet systems that have now been detected
via a variety of techniques. These systems exhibit a range of both
planetary properties and orbital configurations. For those systems
without detected planetary transits, a significant unknown factor is
the orbital inclination. This produces an uncertainty in the mass of
the planets and their related properties, such as atmospheric scale
height. Here we investigate the HD~10180 system which was discovered
using the radial velocity technique. We provide a new orbital solution
for the system which allows for eccentric orbits for all planets. We
show how the inclination of the system affects the mass/radius
properties of the planets and how the detection of phase signatures
may resolve the inclination ambiguity. We finally evaluate the
Habitable Zone properties of the system and show that the g planet
spends 100\% of an eccentric orbit within the Habitable Zone.

\end{abstract}


%%%%%%%%%%%%%%%%%%%%%%%%%%%%%%%%%%%%%%%%%%%%%%%%%%%%%%%%%%%%%%%%%%%%

\section{Introduction}
\label{intro}

Multi-planet systems discoveries have revealed a diversity of system
architectures, many of which significantly diverge from that of our
own Solar System. Many of the recent multi-planet system discoveries
have been made as a result of data from the {\it Kepler} mission, such
as Kepler-62 \citep{bor13}. These systems tend to harbor a mixture of
terrestrial and Neptune-size planets, some of which are in the
Habitable Zone (HZ) of their host star. There have also been several
discoveries of systems with more than four planets that have been
discovered by radial velocity (RV) surveys, such as the 55 Cancri
system \citep{mca04,end12}. The orbital inclination of the planets in
these cases is generally unknown, although they can be constrained
through examination of the dynamical stability of the system (e.g.,
\citet{cor10}).

A multi-planet system of particular interest is the HD~10180 system,
due to the both the relatively large number of planets and their
relatively low masses. There have been various interpretations of the
RV data for this system with respect to the number of planets
present. \citet{lov11} provide a seven planet solution where the
detection of the inner ``b'' planet is considered tentative. A further
solution by \citet{tuo12} demonstrates that the system may harbor nine
planets. Although in both cases the planets likely have low masses,
this depends on the inclination of the system with respect to the
plane of the sky. As the inclination decreases the mass of the planets
increases and thus their physical properties change. The inclination
ambiguity can be resolved using several techniques, such as astrometry
\citep{tuo09} and phase curve analysis
\cite{kan11,kan12a}. Determining the true masses of the planets is a
key factor in determining the significance of their locations within
the stellar HZ \citep{kop14}.

Here we present the results of a new analysis of the HD~10180 system
in which we discuss the orbital parameters, inclination, predicted
phase signatures, and HZ status of the planets. In Section 2 we
provide a new Keplerian orbital solution for the system with eccentric
orbits for all planets. Section 3 investigates the effects of orbital
inclination on planet masses and possible radii. Section 4 discusses
the phase variations of the system in different inclination scenarios
and the detectability of those signatures. In Section 5 we present an
analysis of the system HZ in the context of various inclinations. We
provide concluding remarks in Section 6.

%%%%%%%%%%%%%%%%%%%%%%%%%%%%%%%%%%%%%%%%%%%%%%%%%%%%%%%%%%%%%%%%%%%%

\section{System Configuration}
\label{system}

HD~10180 is a star which is quite similar to solar (G1V) in terms of
its fundamental properties. These are summarized in Table
\ref{stellar}, where the majority of parameters are those provided by
\citet{lov11}. The distance is derived from {\it Hipparcos} parallax
measurements \citep{van07} and the stellar radius is calculated from
the mass-radius relationships determined by \citet{tor10}. As noted by
\citet{lov11}, the activity index shows that HD~10180 is a relatively
inactive star, a property that will be of particular relevance when
discussing the photometry in Section \ref{phase}.

\begin{deluxetable}{lc}
  \tablecaption{\label{stellar} Stellar Parameters$^{(1)}$}
  \tablehead{
    \colhead{Parameter} &
    \colhead{Value}
  }
  \startdata
  $V$                            & 7.33 \\
  $B-V$                          & 0.629 \\
  Distance (pc)$^{(2)}$          & $39.02 \pm 1.1$ \\
  $T_\mathrm{eff}$ (K)           & $5911 \pm 19$ \\
  $\log g$                       & $4.39 \pm 0.03$ \\
  $[$Fe/H$]$ (dex)               & $0.08 \pm 0.01$ \\
  $M_\star$ ($M_\odot$)          & $1.06 \pm 0.05$ \\
  $R_\star$ ($R_\odot$)$^{(3)}$  & $1.109 \pm 0.036$
  \enddata
  \tablenotetext{(1)}{\citet{lov11}}
  \tablenotetext{(2)}{\citet{van07}}
  \tablenotetext{(3)}{\citet{tor10}}
\end{deluxetable}

\begin{figure*}
  \begin{center}
    \includegraphics[width=15.3cm]{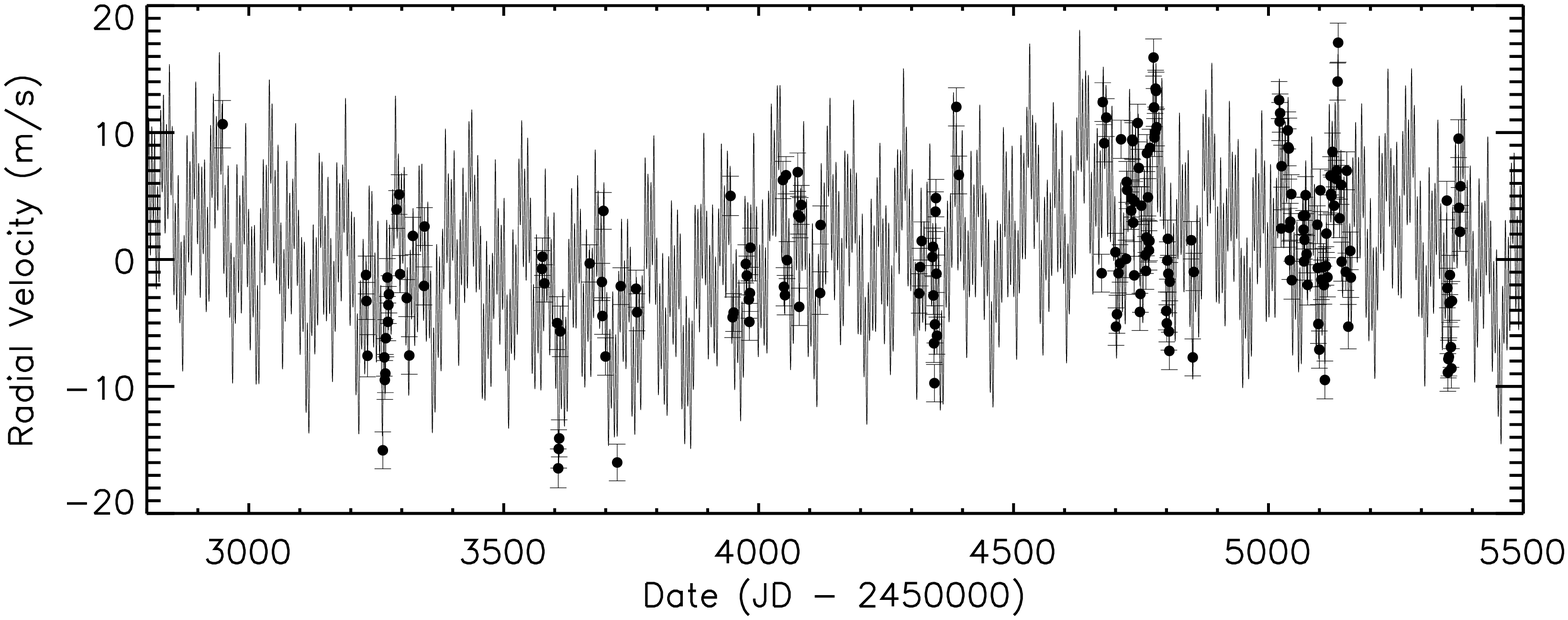} \\
    \includegraphics[width=15.0cm]{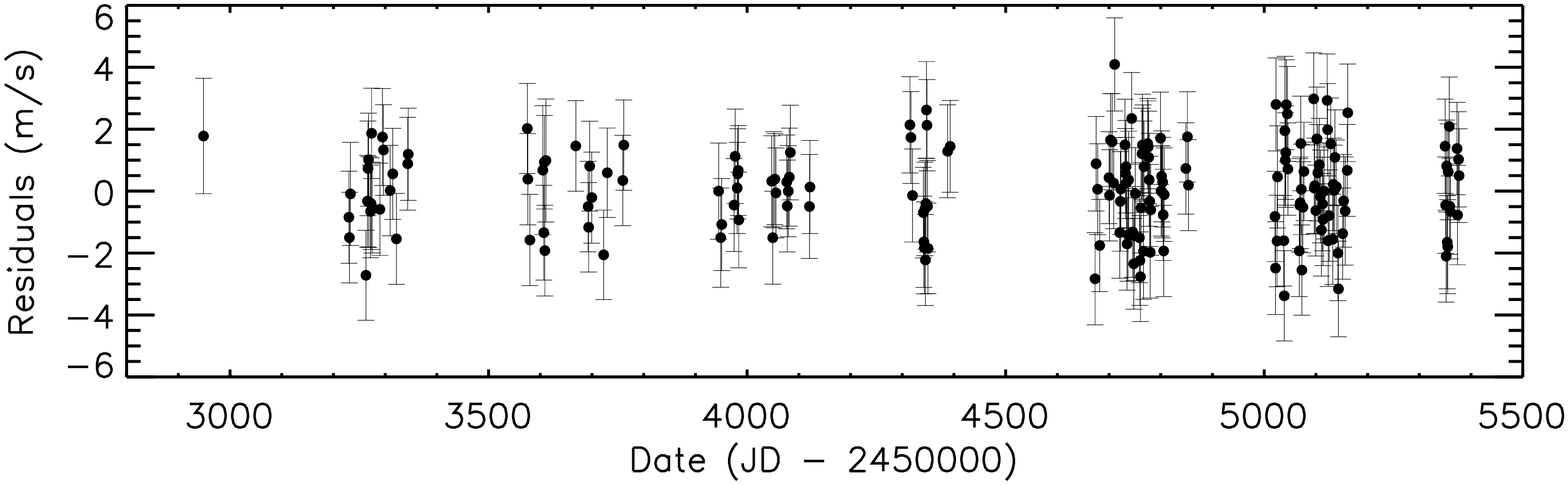}
  \end{center}
  \caption{Top panel: The 190 RV measurements of HD~10180 along with
    the best-fit 6-planet Keplerian solution. The solution, shown in
    Table \ref{planets}, allows the eccentricities of all planets to
    be free parameters. Bottom panel: The RV residuals (observed minus
    computed) from the best-fit model shown above.}
  \label{rvdata}
\end{figure*}

\begin{deluxetable*}{lcccccc}
  \tablecolumns{7}
  \tablewidth{0pc}
  \tablecaption{\label{planets} HD 10180 Planetary Parameters}
  \tablehead{
    \colhead{Parameter} &
    \colhead{c} &
    \colhead{d} &
    \colhead{e} &
    \colhead{f} &
    \colhead{g} &
    \colhead{h}
  }
  \startdata
$P$ (days)
  & $5.75969 \pm 0.00028$  & $16.3570 \pm 0.0038$
  & $49.748 \pm 0.025$     & $122.744 \pm 0.232$
  & $604.67 \pm 10.42$     & $2205.0 \pm 105.9$ \\
$T_p\,^{(1)}$
  & $4001.445 \pm 0.426$  & $4022.119 \pm 1.157$
  & $4006.26 \pm 6.91$    & $4024.67 \pm 11.03$
  & $4002.8 \pm 68.3$     & $3433.4 \pm 393.1$ \\
$e$
  & $0.073 \pm 0.031$      & $0.131 \pm 0.052$
  & $0.051 \pm 0.033$      & $0.119 \pm 0.054$
  & $0.263 \pm 0.152$      & $0.095 \pm 0.086$ \\
$\omega$ (deg)
  & $328 \pm 24$           & $325 \pm 23$
  & $147 \pm 54$           & $327 \pm 27$
  & $327 \pm 59$           & $142 \pm 72$ \\
$K$ (m\,s$^{-1}$)
  & $4.545 \pm 0.154$      & $2.935 \pm 0.173$
  & $4.283 \pm 0.169$      & $2.862 \pm 0.186$
  & $1.754 \pm 0.380$      & $3.117 \pm 0.245$ \\
$M_p \sin i$ ($M_J$)
  & $0.0416 \pm 0.0014$    & $0.0378 \pm 0.0022$
  & $0.0805 \pm 0.0032$    & $0.0722 \pm 0.0047$
  & $0.0732 \pm 0.0138$    & $0.2066 \pm 0.0139$ \\
$a$ (AU)
  & $0.06412 \pm 0.00101$  & $0.12859 \pm 0.00202$
  & $0.2699 \pm 0.0043$    & $0.4929 \pm 0.0078$
  & $1.427 \pm 0.028$      & $3.381 \pm 0.121$
  \enddata
  \tablenotetext{(1)}{BJD -- 2,450,000}
\end{deluxetable*}

The Keplerian orbital solution provided by \citet{lov11} includes
seven planets and forces a circular orbit for several of those
planets, including planet g. Since the semi-amplitude of the RV signal
for the b planet is significantly lower than the others, we performed
our own fit to the RV data to obtain a Keplerian orbital solution in
which all eccentricities were allowed to vary as free parameters. The
RV data were extracted from the VizieR Catalog Service\footnote{\tt
  http://vizier.u-strasbg.fr/}. These consist of 190 measurements
obtained with the HARPS spectrograph at the ESO 3.6m telescope at La
Silla Observatory. We fit the data using the partially linearized,
least-squares fitting procedure described in \citet{wri09} and
estimated parameter uncertainties using the BOOTTRAN bootstrapping
routines described in \citet{wan12}. Our best solution includes six
planets where there is no significant RV trend in the data. We adopted
a slightly larger stellar jitter value than \citet{lov11} of
1.39~m\,s$^{-1}$ which forces the reduced $\chi^2$ value to unity. The
resulting orbital solution is shown in Figure \ref{rvdata} and Table
\ref{planets}. The residuals shown in the bottom panel of Figure
\ref{rvdata} have an RMS scatter of 1.5~m\,s$^{-1}$. The main
differences with the solution by \citet{lov11} are: (1) no planet b,
(2) a significant eccentricity for planet g, and (3) a smaller orbital
period and eccentricity for planet h. Note that \citet{lov11} force
the eccentricity of the g planet to zero since a non-zero eccentricity
produces an almost identical $\chi^2$. Here we consider a non-zero
eccentricity for the g planet since it is consistent with the data and
is relevant to our subsequent habitability discussion in Section
\ref{hab}.

\begin{figure*}
  \begin{center}
    \includegraphics[angle=270,width=15.0cm]{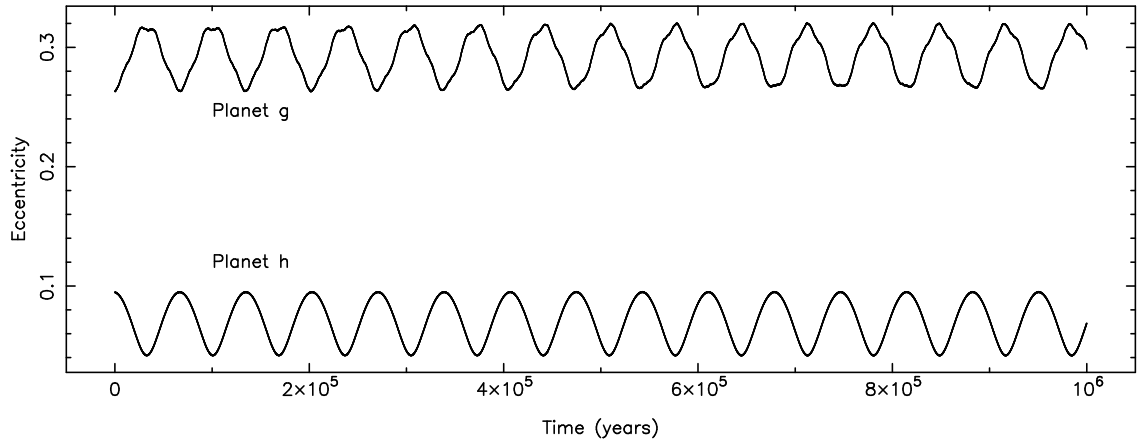}
  \end{center}
  \caption{Dynamical simulations of the HD~10180 system, showing the
    eccentricity oscillations of the g and h planets over a period of
    $10^6$ years. The g planet primarily dynamically interacts with
    the h planet with additional minor perturbations caused by the f
    planet.}
  \label{stability}
\end{figure*}

Since this eccentric solution is different from those previously
published, it is important to establish if it is dynamically sound. To
explore this, we performed dynamical simulations using N-body
integrations with the Mercury Integrator Package, described in more
detail by \citet{cha99}. We adopted the hybrid
symplectic/Bulirsch-Stoer integrator and used a Jacobi coordinate
system. This coordinate system generally provides more accurate
results for multi-planet systems \citep{wis91,wis06} except in cases
of close encounters \citep{cha99}. The integrations were performed for
a simulation of $10^6$ years, in steps of 100 years, starting at the
present epoch.

Our simulations indicate that the orbital configuration presented in
Table \ref{planets} are stable over the $10^6$ year simulation. The
planets do exchange angular momentum through secular oscillations of
their eccentricities, but this remains at a relatively low level. The
two main examples of this are the outermost (g and h) planets. The
eccentricity oscillations for both of these planets are shown in
Figure \ref{stability} for the complete simulation period. The secular
oscillations complete approximately 15 cycles during the $10^6$~year
simulation with a period of $\sim$65,000 years. The range of
eccentricity for the g and h planets are 0.263--0.321 and 0.042--0.095
respectively. The g planet eccentricity oscillations are also
perturbed due to interactions with the f planet. We discuss the
implications of these oscillations further in Section \ref{hab}.

%%%%%%%%%%%%%%%%%%%%%%%%%%%%%%%%%%%%%%%%%%%%%%%%%%%%%%%%%%%%%%%%%%%%

\section{Inclination and Planetary Properties}
\label{inclination}

\begin{figure}
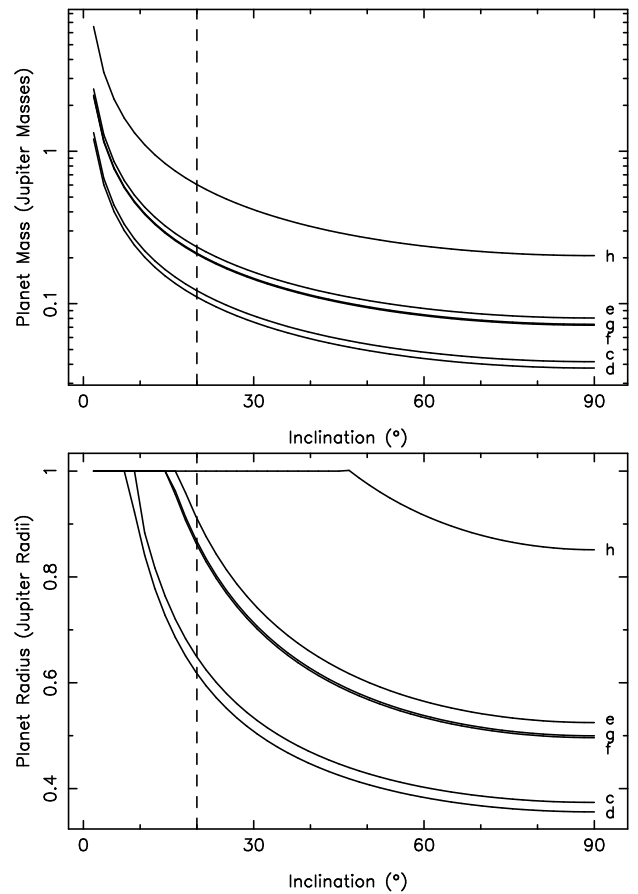

  \begin{center}
    \includegraphics[angle=270,width=8.2cm]{f03a.ps} \\
    \includegraphics[angle=270,width=8.2cm]{f03b.ps}
  \end{center}
  \caption{The dependence of the HD~10180 planetary properties of mass
    (top panel) and radius (bottom panel) on the system inclination,
    where $90\degr$ is an edge-on orientation and $0\degr$ is
    face-on. The vertical lines represent the likely lower inclination
    limit of the system as estimated by \citet{lov11}.}
  \label{incfig}
\end{figure}

There is a well studied relationship between planetary mass and
radius. Early work on mass-radius relationships for gas giants
\citep{for07} and super-Earths \citep{sea07} paved the way for
understanding the results of Kepler discoveries. Kepler planets have
subsequently allowed empirical relations to be developed for low-mass
planets \citep{wei13,wei14}. The nature of radial velocity discoveries
of exoplanet that lack confirmation from other techniques is that it
is only the minimum mass that is known. The true planetary masses
depend on the inclination of the system, from edge-on ($i = 90\degr$)
to face-on ($i = 0\degr$) with respect to the plane of the sky.

In Figure \ref{incfig} we show the increase in the HD~10180 planet
masses as a function of orbital inclination. The f and g planets have
similar masses and so their lines in the plots are almost
indistinguishable. The dynamical analysis of the HD~10180 system by
\citet{lov11} shows that the system is still stable for $i = 30\degr$
but not for $i = 10\degr$, concluding that an instability transition
occurs around $i \sim 20\degr$. We tested this instability transition
by repeating our stability analysis described in Section \ref{system}
for a variety of inclinations. We confirm that the system becomes
unstable at $i \sim 20\degr$ (shown as a vertical dashed line in
Figure \ref{incfig}) with the ejection of the d planet, but the g
planet remains stable despite the higher eccentricity. The range of g
planet eccentricities described in Section \ref{system} remains the
same until instability occurs. The bottom panel of Figure \ref{incfig}
shows the corresponding change in planet radius using the simple
mass-radius relationship of \citet{kan12b} which assumes an
approximate Jupiter radius for masses larger than 0.3 Jupiter
masses. We discuss the implications of these mass/radius increases in
the following sections.

%%%%%%%%%%%%%%%%%%%%%%%%%%%%%%%%%%%%%%%%%%%%%%%%%%%%%%%%%%%%%%%%%%%%

\section{Phase Variations}
\label{phase}

\begin{figure}
  \includegraphics[angle=270,width=8.2cm]{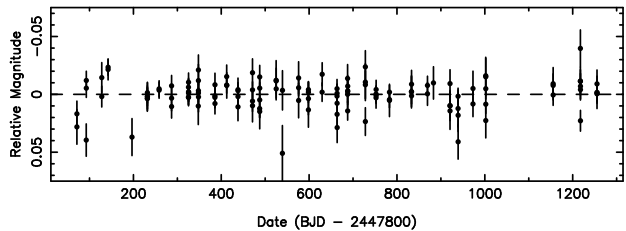}
  \caption{Photometry of HD~10180 from the {\it Hipparcos} mission
    which shows photometric stability at the 1\% level.}
  \label{hipparcos}
\end{figure}

A further means through which to constrain the inclination of the
system is by the detection of phase variations. This technique was
described in detail by \citet{kan12a} where the amplitude of the phase
variations depend on the planetary properties which vary with
inclination (see Section \ref{inclination}). One aspect of the system
that affects the ability to detect such phase signatures is the
activity of the star. HD~10180 is known to be a relatively inactive
star with a mean activity index of $\log R_{\mathrm{HK}}^{'} = -5.00$
\citep{lov11}. We used publicly available data from the {\it
  Hipparcos} satellite to search for low-frequency photometric
variations of HD~10180. {\it Hipparcos} acquired a total of 125
measurements spanning a period of 1184 days during the course of its
three-year mission \citep{per97,van07}. These data are shown in Figure
\ref{hipparcos}. The 1$\sigma$ RMS scatter of the 125 HD~10180
measurements is 0.013 mag, while the mean of the measurement
uncertainties is 0.011. Thus it is consistent with a photometrically
stable star at the 1\% level. A Fourier analysis of the {\it
  Hipparcos} data do not reveal any significant periodic signatures
and indeed the data Nyquist frequency of 0.0528~days$^{-1}$ is
slightly above the predicted period of the stellar rotation ($\sim
24$~days). The {\it Hipparcos} data sampling is therefore unlikely to
detected stellar rotation variability.

\begin{figure*}
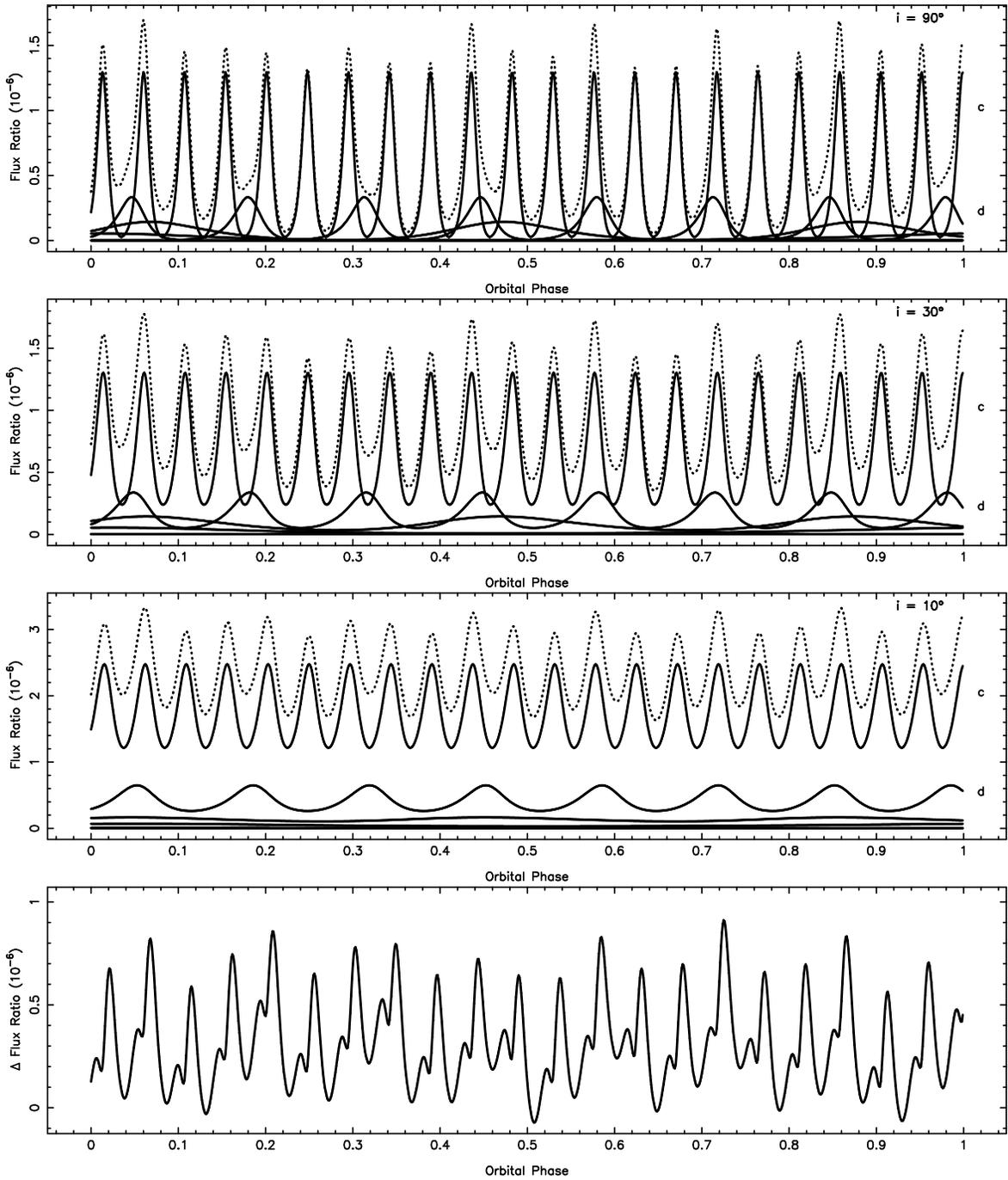

  \begin{center}
    \includegraphics[angle=270,width=15.5cm]{f05a.ps} \\
    \includegraphics[angle=270,width=15.5cm]{f05b.ps} \\
    \includegraphics[angle=270,width=15.5cm]{f05c.ps} \\
    \includegraphics[angle=270,width=15.5cm]{f05d.ps}
  \end{center}
  \caption{The flux variations of the HD~10180 system due to the phase
    variations of the planets as they orbit the host star. We consider
    three inclinations of the plane of the system with respect to the
    plane of the sky; $i = 90\degr$ or edge-on (top panel), $i =
    30\degr$ (second panel), and $i = 10\degr$ (third panel). In each
    panel the phase variations of individual planets are shown as
    solid lines phased on the orbital period of the outer planet, the
    combined signature of all six planets is shown as a dotted line,
    and the phase variations of c and d are labeled. The bottom panel
    shows the difference in flux variations between the $i = 10\degr$
    and $i = 90\degr$ scenarios after the baseline flux has been
    removed.}
  \label{phasefig}
\end{figure*}

Although it is advantageous that the star is relatively quiet, the
phase variations occur on a much lower level. We calculate the
predicted phase variations of the system by adopting the formalism of
\citet{kan10}. This formalism accounts for planetary size, orbital
eccentricity, and the variation of geometric albedo with separation
from the host star. The flux ratio of the planet to the host star is
given by
\begin{equation}
  \epsilon(\alpha,\lambda) \equiv
  \frac{f_p(\alpha,\lambda)}{f_\star(\lambda)} = A_g(\lambda)
  g(\alpha,\lambda) \frac{R_p^2}{r^2}
  \label{fluxratio}
\end{equation}
where $A_g$ is the geometric albedo, $g(\alpha,\lambda)$ is the phase
function, $R_p$ is the planetary radius, and $r$ is the star--planet
separation. The resulting flux variations of the system are shown in
the top three panels of Figure \ref{phasefig} where we have calculated
the variations for system inclinations of $90\degr$ (edge-on),
$30\degr$, and $10\degr$. In each panel the phase flux variations due
to the individual planets are shown as solid lines and the combined
variations are indicated by the dotted line. These are shown for one
complete orbital phase of the outer planet. As described in Section
\ref{inclination}, the inclination must be larger than $10\degr$ in
order to retain a stable orbital configuration for the system.

The phase variations of the c and d planets are also labeled on the
right of each panel in Figure \ref{phasefig}. The peak flux variations
are dominated by the inner (c) planet for each inclination. The planet
has a calculated radius of 0.37 and 0.92 Jupiter radii for
inclinations of $90\degr$ and $10\degr$ respectively. As shown by
\citet{kan11}, the effect of decreasing the inclination is to remove
the time variability of the phase function resulting in flux
variations caused exclusively by orbital eccentricity. This can be
particularly seen for the d planet in the bottom panel which retains a
photometrically variable signature due to its eccentricity of 0.131
(see Table \ref{planets}). The amplitude of the total variations
remains very similar with decreasing inclination due to the
compensation of the increased planetary radii. However, the constantly
visible illumination of the c and d planets for lower inclinations
significantly raises the baseline of the planetary reflected light
received. The effect of this is to raise to signal-to-noise of the
variations which makes their detection more accessible. A possible
method to discriminate between the baseline flux from the planet and
the stellar flux is through polarized light. The motion of the
planet(s) will produce a polarization signature distinct from the
stellar flux due to the scattering of light from the planetary
atmospheres \citep{ber11,wik09}. However, for Keplerian orbits an
additional distinction is available via the difference in phase
between the times of periastron and maximum phase variations (phase
angle zero). The bottom panel of Figure \ref{phasefig} shows the
difference between the combined flux variations of $i = 10\degr$ and
$i = 90\degr$ after the minimum flux (baseline) has been removed. The
$58\degr$ separation of the periastron passage of the c planet from
the zero phase angle produces a difference in phase signature of
amplitude similar to the individual inclination scenarios. In
practice, most systems will have an even larger separation which will
aid in this distinction, the amplitude of which will depend on the
orbital eccentricities. Resolving this degeneracy will greatly aid in
disentangling the components of the phase signature and thus the
inclination of the system. The total flux variations are of amplitude
several parts per million (ppm) and thus close to the photometric
precision achieved by the {\it Kepler} mission for the brightest stars
monitored. This technique could therefore be used to rule out low
system inclinations and/or high albedos for multi-planet systems,
particularly with data from future missions such as the James Webb
Space telescope (JWST) and the Transiting Exoplanet Survey Satellite
(TESS).

%%%%%%%%%%%%%%%%%%%%%%%%%%%%%%%%%%%%%%%%%%%%%%%%%%%%%%%%%%%%%%%%%%%%

\section{Habitability of the \lowercase{g} Planet}
\label{hab}

\begin{figure*}
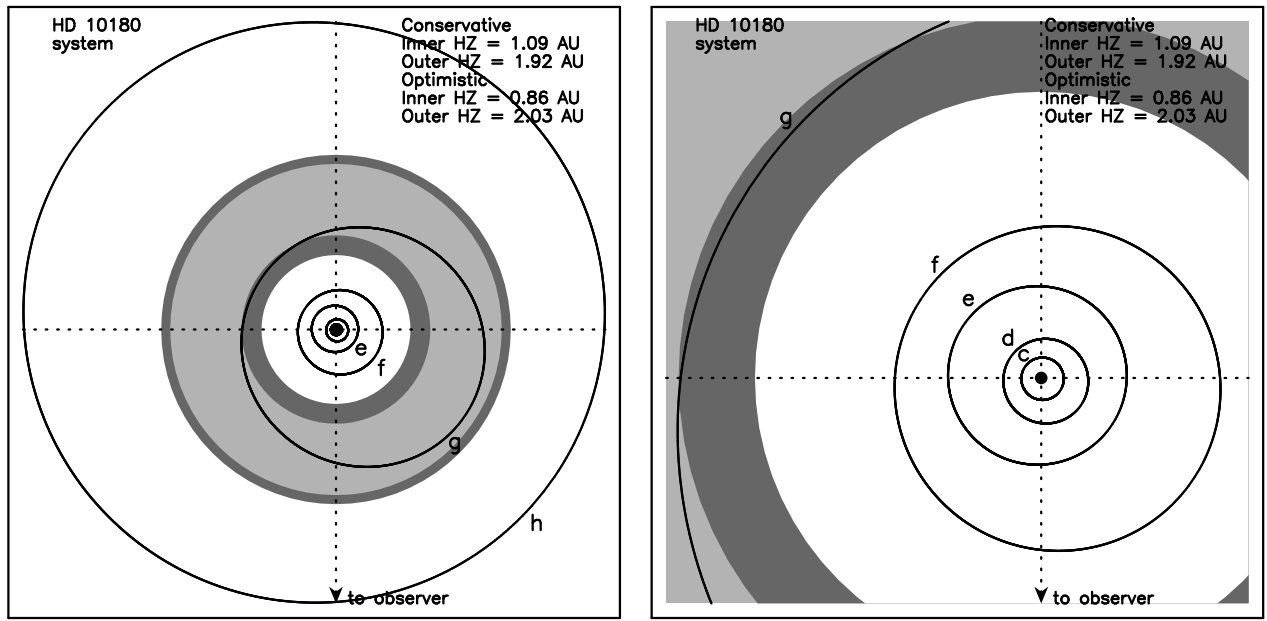

  \begin{center}
    \begin{tabular}{cc}
      \includegraphics[angle=270,width=8.2cm]{f06a.ps} &
      \includegraphics[angle=270,width=8.2cm]{f06b.ps}
    \end{tabular}
  \end{center}
  \caption{A top-down view of the HD~10180 system showing the extent
    of the HZ calculated using the stellar parameters of Table
    \ref{stellar}. The conservative HZ is shown as light-gray and
    optimistic extension to the HZ is shown as dark-gray. The revised
    Keplerian orbits of the planets from Table \ref{planets} are
    overlaid. Left Panel: The full HZ of the system with the outer 5
    planets. Right panel: A zoom-in of the system showing the orbits
    of the inner 5 planets and the orbital path of the g planet into
    the optimistic HZ.}
  \label{hzfig}
\end{figure*}

With the revised orbital solution of Section \ref{system} and
inclination effects described in Sections \ref{inclination} and
\ref{phase}, we finally investigate the HZ of the system. The
empirically derived HZ boundaries of \citet{kas93} have recently been
replaced by new calculations by \citet{kop13,kop14}. We use these
calculations to provide new estimates for the HZ in the HD~10180 by
adopting the definitions of ``conservative'' and ``optimistic'' HZ
models described by \citet{kan13}. These definitions use different
boundaries for the HZ based on assumptions regarding the amount of
time that Venus and Mars were able to retain liquid water on their
surfaces \citep{kop13}. \citet{kan14} showed the extent to which HZ
boundaries depend on stellar parameter uncertainties, although the
parameters in Table \ref{stellar} are sufficiently well known that the
HZ boundary uncertainties are negligible. HZ calculations for all
known exoplanetary systems are available using the same methodology
through the Habitable Zone Gallery \citep{kan12b}.

Figure \ref{hzfig} shows two top-down views of the HD~10180 system,
one zoomed out to include the outer planet (left panel) and the other
zoomed in to show the inner planets (right panel). In each panel, the
HZ is depicted by the shaded region where the light gray represents
the conservative HZ and the dark gray is the extension to the HZ with
optimistic calculations. Of greatest interest in this figure from a HZ
perspective is the g planet. The planet remains in the conservative HZ
for 89\% of the duration of one orbit with the remaining 11\% within
the optimistic HZ. The orbital path through the optimistic HZ occurs
during the periastron passage and is clearly shown in the right panel
of Figure \ref{hzfig}. The orbital stability analysis from Section
\ref{system} showed that the eccentricity of the planet can oscillate
to a value as high as $\sim$0.32 over $\sim$65,000 year timescales. We
recomputed our HZ calculations using this maximum eccentricity and
found that the planet remains 100\% in the HZ; 83\% in the
conservative HZ and 17\% in the optimistic HZ. Thus interactions with
the other planets do not perturb the HZ status of the planet. Note
that increasing the eccentricity by the 1$\sigma$ uncertainty shown in
Table \ref{planets} ($e = 0.415$) results in the planet passing
slightly interior to the optimistic HZ during periastron passage, but
it remains in the HZ for 94\% of the orbital period. Planets in
eccentric orbits which spend only part of their orbit in the HZ have
been previously investigated \citep{kan12c,wil02}. These studies show
that habitability is not necessarily ruled out depending on the
efficiency of the planetary atmosphere in redistributing the variable
energy from stellar insolation during the orbit.

For an inclination of $90\degr$, the mass of the g planet is 0.0732
Jupiter mass or 23.3 Earth masses. The estimated radius according to
Figure \ref{incfig} is 0.5 Jupiter radii. An inclination of $30\degr$
raises these values to 0.1464~$M_J$, 46.5~$M_\oplus$, and 0.71~$R_J$
respectively. Adjusting the inclination to the stability limit of
$10\degr$ further raises the values to 0.422~$M_J$, 134.0~$M_\oplus$,
and 1.0~$R_J$ respectively. For any value of inclination, the mass of
the g planet is far above the threshold where the planet is likely to
have a purely rocky composition \citep{mar14}. Thus the prospects for
habitability for the g planet lies within a moon system which the
planet may harbor. Searches for exomoons around such HZ giant planets
have been undertaken \citep{kip13} but have not yet yielded positive
results. Recent studies of exomoon habitability have shown that there
are a variety of factors which add to the total energy budget
including flux from the planet and tidal effects
\citep{hel12,hin13}. Although these additional factors are usually
negligible compared with the flux from the host star, they may be
sufficient in this case to render the g planet moons devoid of surface
liquid water considering the planet already moves interior to the
conservative HZ.

%%%%%%%%%%%%%%%%%%%%%%%%%%%%%%%%%%%%%%%%%%%%%%%%%%%%%%%%%%%%%%%%%%%%

\section{Conclusions}

Multi-planet exoplanetary systems offer exceptional opportunities for
system characterization, such as constraining the orbital inclinations
and eccentricities based on stability simulations. These systems are
particularly interesting when one or more of the planets occupy the HZ
of their host star. The HD~10180 is just such a system, with a range
of planetary masses and at least one planet within the HZ. We have
presented a new orbital solution which allows the g planet to have
significant orbital eccentricity whilst preserving the stability of
the system. We have quantitatively shown how the properties of the
planets alter depending on the inclination of the system. These
properties in turn change the predicted phase signatures of the system
and provides a method through which future observations could resolve
the inclination ambiguity.

An important consideration for HZ planets is the orbital eccentricity
since the variable stellar insolation can greatly effect the
habitability. The revised orbital solution presented here allows for
an eccentric orbit for the only HZ planet in the system. The HZ
calculations described here show that the g planet spends most of a
complete orbital period within the conservative HZ and moves into the
interior optimistic HZ for the remaining time. The mass of the planet
is high enough that surface liquid water is only possible on any moons
the planet may possess. The topic of exomoons is one of increasing
study and detection efforts and is thus likely to yield positive
detections in the near future. As next-generation instrumentation is
developed for exoplanetary studies, it is important to identify the
best HZ targets orbiting bright host stars such as the one described
here.

%%%%%%%%%%%%%%%%%%%%%%%%%%%%%%%%%%%%%%%%%%%%%%%%%%%%%%%%%%%%%%%%%%%%

\section*{Acknowledgements}

The authors would like to thank the anonymous referee, whose comments
greatly improved the quality of the paper. This research has made use
of the following online resources: the Exoplanet Orbit Database and
the Exoplanet Data Explorer at exoplanets.org, the Habitable Zone
Gallery at hzgallery.org, and the VizieR catalog access tool, CDS,
Strasbourg, France.

%%%%%%%%%%%%%%%%%%%%%%%%%%%%%%%%%%%%%%%%%%%%%%%%%%%%%%%%%%%%%%%%%%%%

\end{document}